\documentclass[aps,prl,twocolumn,groupedaddress]{revtex4}
\pdfoutput=1
\usepackage{graphicx}

\begin{document}


\title{On the use of black hole binaries as probes of local dark energy properties}


\author{Jonas Enander}
\email{enander@fysik.su.se}
\author{Edvard M\"ortsell}
\email{edvard@fysik.su.se}
\affiliation{The Oskar Klein Centre for Cosmoparticle Physics, 
Department of Physics, Stockholm University, AlbaNova University
Center, S--106 91 Stockholm, Sweden}


\date{\today}

\begin{abstract}
Accretion of dark energy onto black holes will take place when dark
energy is not a cosmological constant. It has been proposed that the
time evolution of the mass of the black holes in binary systems due
to dark energy accretion could be detectable by gravitational
radiation. This would make it possible to use observations of black
hole binaries to measure local dark energy properties, e.g., to
determine the sign of $1+w$ where $w$ is the dark energy equation of
state. In this Letter we show that such measurements are unfeasible due
to the low accretion rates.
\end{abstract}


\maketitle

\section{1. Introduction}

Experimental evidence shows that the universe currently undergoes an
accelerated expansion driven by an unknown form of energy, dubbed dark
energy. This dominant energy component can -- at least
phenomenologically -- be described as a perfect fluid with density
$\rho_{de}$ and a (possibly time dependent) equation of state
$w\equiv p_{de}/\left(\rho_{de}c^{2}\right)$, where $p_{de}$ is the dark energy
pressure. Current measurements are consistent with the dark energy
density being approximately 70\,\% of the critical density of the
universe, $\rho_{de}\sim 10^{-29}\,\mbox{g cm}^{-3}$, and the equation
of state being close to $w=-1$, i.e., the value corresponding to a
cosmological constant. Since the density is very low and we do not
expect dark energy to cluster to any large extent, all the information
we have on dark energy is through its impact on the largest scales of
the universe via the expansion rate of the universe. One of the
foremost experimental and theoretical tasks in cosmology is to pin
down the equation of state $w$ of the dark energy, specifically to
constrain or detect any deviations from the cosmological constant
value of $w=-1$.

It has been proposed that dark energy can be detected also on smaller
scales \cite{mersini,wang}. Any such local measurements would
constitute a powerful confirmation of the existence of dark energy. In
\citet{mersini}, the possibility to measure local properties of dark
energy using gravitational radiation from binary systems of
supermassive black holes is investigated. The basic idea is that
accretion of dark energy by black holes will lead to a mass change
when dark energy is not a cosmological constant. The mass accretion
would affect the dynamics of the system and in turn alter the waveform
of the gravitational radiation produced by the binaries during the
inspiraling phase. Measuring this waveform, through LIGO or LISA,
would then make it possible to constrain the values of
$w$ and $\rho_{de}$. Alternatively, the change in orbital period of the system could
be measured using electromagnetic radiation, if present. In this Letter
we claim that this effect is negligible and that this form of local
measurement of $\rho_{de}$ and $w$ is not possible.

In Section 2 of the Letter, we compare the mass accretion due to
the dark energy with that of the interstellar medium and dark
matter. In Section 3 we look at the radial separation needed for
dark energy accretion to dominate over the effect over gravitational
radiation in a binary system of two supermassive black holes. In both
cases we show that dark energy accretion cannot be considered a
measurable effect, even for a binary system. In Section 4 we
consider two concrete examples to validate this conclusion.

\section{2. Mass accretion}
The formula for a spherically symmetric, adiabatic, steady-flow gas
accretion onto a star at rest with respect to the gas was developed by
Bondi in 1952 \cite{bondi}. Since the properties of the mass accretion
rate is determined by the gravitational field far away from the
accreting object, it can be used to estimate the accretion rate for a
black hole with respect to the interstellar medium. For a medium with
mean free path length smaller than the accretion radius, this can be
expressed as \cite{frank}
\begin{eqnarray}
\dot{m}\cong&&1.4\times10^{11}\left(\frac{m}{m_{\odot}}\right)^{2}\nonumber\\
&&\times\left(\frac{\rho\left(\infty\right)}{10^{-24}\,\mbox{g
cm}^{-3}}\right)\left(\frac{c_{s}\left(\infty\right)}{10\,\mbox{km
s}^{-1}}\right)^{-3}\mbox{g s}^{-1}\, .
\label{eq:nonrelacc}
\end{eqnarray}
Here $\rho\left(\infty\right)$ is the density of the surrounding
medium at infinity and $c_{s}\left(\infty\right)$ is the sound speed
of the medium at infinity. For a supermassive black hole with mass
$10^{8}m_{\odot}$, this leads to an accretion rate of order
$\dot{m}\sim 10^{27}\,\mbox{g s}^{-1}\sim 20\,m_{\odot}\mbox{yr}^{-1}$ in the case of a constant gas supply.

In the case of an object moving with velocity $v$ with respect to the
gas, which is at rest far away from the hole, the accretion rate
becomes, for $v^{2}\gg c_{s}^{2}$ (see \cite{bondi})
\begin{eqnarray}
\dot{m}\cong&&5.2\times10^{6}\left(\frac{m}{m_{\odot}}\right)^{2}\nonumber\\
&&\times\left(\frac{\rho\left(\infty\right)}{10^{-24}\,\mbox{g
cm}^{-3}}\right)\left(\frac{v}{300\,\mbox{km
s}^{-1}}\right)^{-3}\mbox{g s}^{-1}\, .
\label{eq:nonrelaccmovobj}
\end{eqnarray}
The reference velocity $300\,\mbox{km s}^{-1}$ is established from the
movement of a black hole in the gas of a merging galaxy (see
\cite{struck}). The accretion rate is thus diminished by a factor
$10^{5}$.

For cold dark matter, the accretion rate is suppressed by a factor
$\left[c_s\left(\infty\right)/c\right]^{2}$ since it is collissionless
\cite{shapiro}. For dark matter densities in the galaxy core of
$\rho\sim10^{-25}\,\mbox{g cm}^{-3}$ and sound speed
$c_s=100\,\mbox{km s}^{-1}$ (see \cite{ostriker}), this yields an
accretion rate of $\dot{m}\sim10^{16}\,\mbox{g s}^{-1}\sim
10^{-10}\,m_{\odot}\mbox{yr}^{-1}$. In the case of a binary system of
two supermassive black holes, we thus expect the mass accretion to be
dominated by interstellar gas with a bulk flow similar to the
velocities of the black holes.

We now turn to the question whether dark energy accretion is
non-negligible. A relativistic formula for the accretion flow onto
compact objects was developed by \citet{michel}. The relativistic
treatment gives corrections of order unity to the accretion flow onto
a black hole \cite{shapiro}. A general treatment of black hole
accretion of a perfect fluid was presented in \citet{babichev}. In
particular, the dependence on the sign of the accretion of the
equation of state of the fluid was investigated. For dark energy,
taken to be in the form of a perfect fluid, the result is
\begin{equation}
\dot{m}=\frac{4\pi AG^{2}\rho_{de}\left(\infty\right)m^{2}\left(1+w\right)}{c^{3}}\, .
\label{eq:relacc}
\end{equation}
The constant $A$ is determined by demanding continuity of the flow
from infinity to the black hole horizon. It depends on the equation of
state and is of order unity. In the rest of the Letter we use $A=1$. $\rho_{de}\left(\infty\right)$ is the
dark energy density at infinity. The equation of state determines the
sign of $\dot{m}$, which lies at the heart of the proposal to do local
measurements of $w$ using binary black hole systems. In passing, we
note that Eq.~(\ref{eq:relacc}) is also valid for an ultrarelativistic
gas in which case $w=1/3$. Since $\rho_{cmb}\lesssim
10^{-4}\rho_{de}$, accretion of cosmic microwave background photons
will be negligible compared to dark energy accretion, unless the dark
energy equation of state is extremely close to unity.

Putting Eq.~(\ref{eq:relacc}) in a form similar to
Eq.~(\ref{eq:nonrelacc}) yields
\begin{eqnarray}
\dot{m}\cong&&6.7\times10^{-8}(1+w)\nonumber\\
&&\times\left(\frac{m}{m_{\odot}}\right)^{2}\left(\frac{\rho_{de}
\left(\infty\right)}{10^{-29}\,\mbox{g
cm}^{-3}}\right)\mbox{g s}^{-1}\, .
\label{eq:relaccnormalized}
\end{eqnarray}
For a supermassive black hole with mass $10^{8}m_{\odot}$ and a dark
energy density close to the critical energy density, the accretion
rate is of the order $\dot{m}\sim10^{8}\,\mbox{g
s}^{-1}\sim10^{-18}\,m_{\odot}\mbox{yr}^{-1}$. Comparing the accretion rate of dark energy with that
from the interstellar medium and dark matter, we see that the possibility of
using gravitational radiation observation to measure the sign of $\dot{m}$ due to
dark energy accretion is effectively zero. We have not studied the possible
numerical results of the investigations carried out in \cite{wang}, which describes
the effect of dark energy on the quasinormal modes associated with the ringdown
phase of the black hole resulting from the coalescence. In principle, however,
this effect is the result of the mass accretion given in Eq.~(\ref{eq:relacc}), and as we have seen, the accretion rate is several orders of magnitude lower than that from the interstellar medium and dark matter. We therefore anticipate
that the effect of dark energy accretion on the quasinormal modes will also be
negligible.

Assuming that dark energy accretion is the dominating mechanism
for black hole mass change, we can solve Eq.~(\ref{eq:relacc}) to obtain
\begin{equation}
m(t)= \frac{m_0}{1-\frac{t}{\tau}}\, ,
\label{eq:mt}
\end{equation}
where
\begin{equation}
\tau=\frac{3\times 10^{40}}{1+w}\left(\frac{\rho_{de}
\left(\infty\right)}{10^{-29}\,\mbox{g cm}^{-3}}\right)^{-1}
\left(\frac{m_0}{m_{\odot}}\right)^{-1}\,\mbox{s}\, .
\label{eq:tau}
\end{equation}
For a $m=10^8 m_\odot$ black hole, $|1+w|\sim 0.1$ and a dark energy
density close to the critical, this corresponds to $\tau = 10^{26}$ years or approximately $10^{16}$ times the age of the
universe.

Although miniscule, we note that dark energy accretion will be larger
than the mass loss due to Hawking radiation for black holes masses
larger than $m\sim10^{-9}m_{\odot}$, i.e., a tenth of the mass of
the moon.

\section{3. Radial change}
In principle, mass accretion in a binary black hole system would give
an imprint on the gravitational waves produced in the system. We make the assumption that the
dark energy does not carry any angular momentum, and that the angular momentum transfer during dark energy accretion therefore is zero. The reason for this is twofold. First, when dark energy is not a cosmological constant, it does admit spatial and temporal variations. These variations are, however, on such large scales that it is hard to see how they would effectively transfer angular momentum to a binary system. Unless the dark energy in the vicinity of the binary system has an angular velocity orders of magnitude larger than that of the binary system, the order of magnitude of the numerical results of this Letter will not be affected.  Secondly, we want to isolate the effect of the mass accretion on the radial separation, which is done under the assumption of zero angular momentum transfer. To affect the waveform, the radial change induced by the
mass accretion under conservation of angular momentum would have to be
of the same order of magnitude as the radial change induced from the
back-reaction of gravitational radiation. In the following we will see
that the radial separations needed for the two effects, dark energy
accretion and gravitational radiation, to be of the same order of
magnitude is on the scale of parsecs. At these separations however,
both effects are negligible. In the following we take $w\neq-1$ so
that $|1+w|\sim 0.1$ and $\rho_{de}\sim 10^{-29}\,\mbox{g cm}^{-3}$
when investigating the effects of dark energy accretion. Since we are
only interested in order of magnitude estimates of the effect, it does
not matter if $1+w$ is positive or negative.

We will consider a binary system with two black holes of equal mass
$m$ during the inspiraling phase in a semi-circular orbit. For
separations much larger than the Schwarzschild radii of the black
holes, $r\gg r_{s}=2Gm/c^{2}$, we can describe a binary black hole
system using Keplerian dynamics. Such a system will produce
gravitational radiation that carry energy away from the system at a
rate
\begin{equation}
P_{gw}=-\frac{64}{5}\frac{G^{4}}{c^{5}}\frac{m^{5}}{r^{5}}\, .
\label{eq:powergrav}
\end{equation}
Equating the orbital energy loss with the power output,
$dE_{orbital}/dt=P_{gw}$, gives the rate of change of the radial
separation $r$ of the two masses:
\begin{equation}
\dot{r}=-\frac{128}{5}\frac{G^{3}}{c^{5}}\frac{m^{3}}{r^{3}}\, .
\label{eq:radgrav}
\end{equation}
To see the effect of dark energy accretion on the radial separation, we
use the fact that angular momentum must be conserved in the
process. Putting $\dot{J}=0$ and using Kepler's law gives
\begin{equation}
\dot{r}=-\frac{2\dot{m}r}{m}\, .
\label{eq:radde}
\end{equation}
The change in orbital energy due to
dark energy accretion is 
\begin{equation}
P_{acc}=-\frac{2G\dot{m}m}{r}\, . 
\label{eq:powerde}
\end{equation}
To consider the joint effect of gravitational radiation and dark
energy accretion we put
\begin{equation}
\frac{dE_{orbital}}{dt}=P_{gw}+P_{acc}\, ,
\label{eq:powerloss}
\end{equation}
which gives 
\begin{equation}
\dot{r}=-\frac{128}{5}\frac{G^{3}}{c^{5}}\frac{m^{3}}{r^{3}}-\frac{2\dot{m}r}{m}\, ,
\label{eq:radtot}
\end{equation}
that is, the total rate of change is equal to the sum of the
individual contributions. Eq.~(\ref{eq:radtot}) is valid as long
as $\dot{m}\ll m\omega$, where $\omega$ is the angular frequency of
the system given by,
\begin{equation}
\omega=\left(\frac{Gm_{tot}}{r^3}\right)^{1/2}\, ,
\label{eq:omega}
\end{equation}
where $m_{tot}$ is the total mass of the binary system. If the mass
accretion rate is of the same order as $m\omega$, we would have to
change the derivation of the gravitational waveform (since the
waveform depends on the second time derivative of the quadrupole
moment) and thus in turn alter the formula for the energy radiated
away from the system.  Eq.~(\ref{eq:radtot}) is different from
the formula for radial change derived in
\citet{mersini}, the reason being that they include the rest energy 
in the orbital energy and does not include $P_{acc}$ in the time
derivative of the orbital energy in Eq.~(\ref{eq:powerloss}). Although
we do not agree on the formula for the time derivatives of the orbital
energy and the separation, these differences do not affect the main
conclusions of this Letter.

Combining Eqs.~(\ref{eq:relacc}) and (\ref{eq:radtot}), we see that
the dark energy accretion term is proportional to $m$ whereas the
gravitational radiation term is proportional to $m^{3}$. The effect of
dark energy accretion would therefore have a larger impact for black
holes with small masses. On the other hand, the frequency of the
gravitational wave is proportional to the square root of the mass
during the inspiraling phase, and in order for the gravitational
radiation to be detectable with LISA or LIGO, we need a system of
supermassive black holes. LISA \footnote{\tt http://lisa.nasa.gov}
will be sensitive in the frequency range
$0.03\,\mbox{mHz}-0.1\,\mbox{Hz}$ whereas LIGO has its maximum
sensitivity around $100\,\mbox{Hz}$ \cite{ligo}. 

\begin{figure}
\includegraphics[angle=0,width=.5\textwidth]{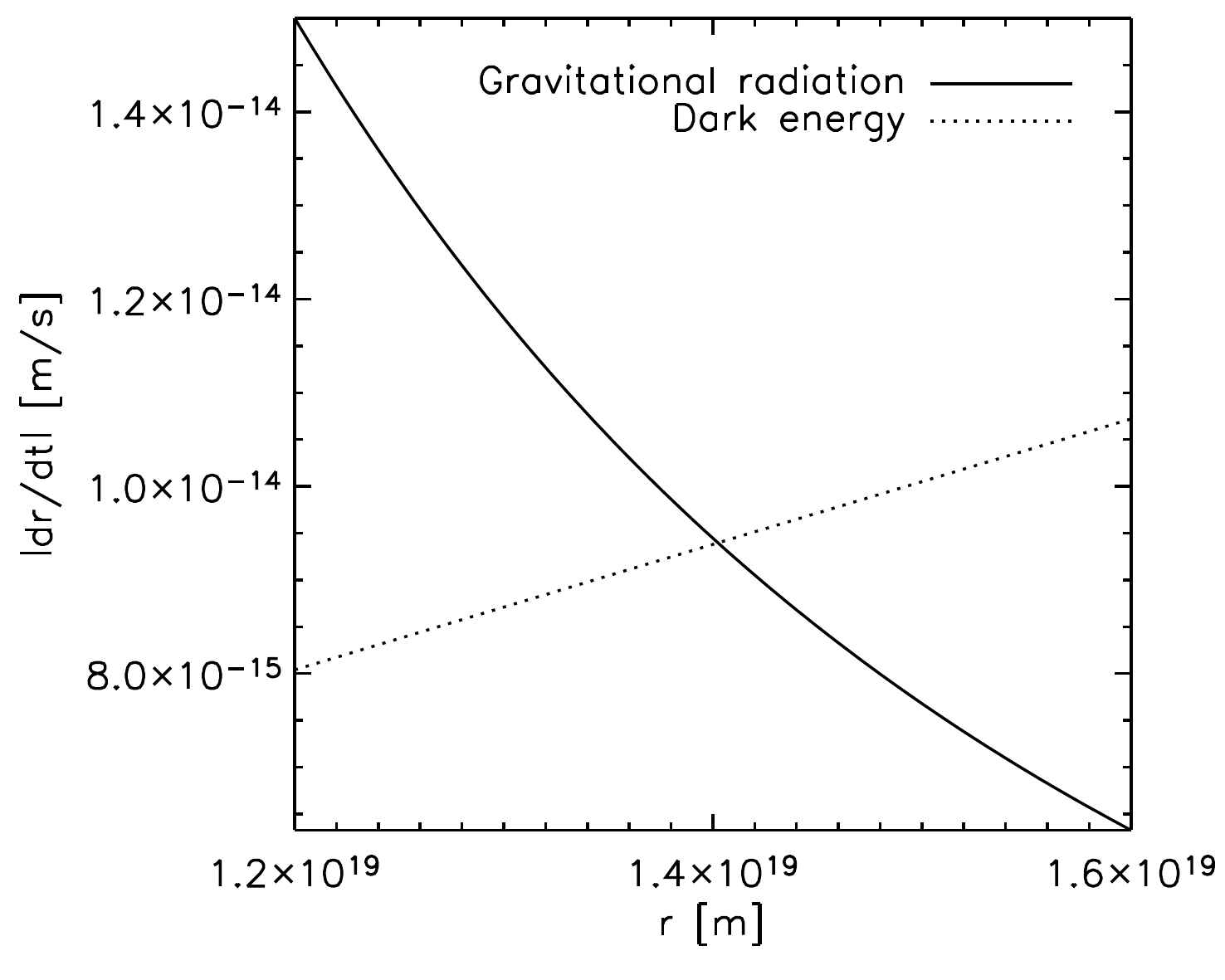}
\caption{\label{fig:billionmass} Radial change as a function of 
separation induced by gravitational radiation and dark energy
accretion for a binary system with equal masses $10^8m_{\odot}$. In
order for dark energy accretion to dominate over gravitational
radiation effects, we need the separation to be larger than
$r\sim10^{19}\,\mbox{m}\sim 1/3\,\mbox{kpc}$. At such large distances,
both effects are negligible.}
\end{figure}

In Fig.~(\ref{fig:billionmass}), we plot the radial change due to both
terms of Eq.~(\ref{eq:radtot}) with respect to the radial separation
in the region where they overlap. The system consists of two black
holes with equal masses $m=10^{8}m_{\odot}$. We see that the effects
are of the same order of magnitude at separations $r\sim10^{19}$~m
with $\left|\dot{r}\right|\sim10^{-14}\,\mbox{m s}^{-1}$.  In general, for a system
of two black holes with equal masses $m$, the separation needs to be at least
$r\sim 10^{15}\left(m/m_{\odot}\right)^{1/2}$~m in order for dark
energy to dominate over gravitational radiation, corresponding to
$\left|\dot{r}\right|\sim10^{-26}\left(m/m_{\odot}\right)^{3/2}\,\mbox{m s}^{-1}$
or $\left|\dot{r}/r\right|=10^{-41}\left(m/m_{\odot}\right)\,\mbox{s}^{-1}$. The
gravitational radiation angular frequency is given by
$10^{-12}\left(m/m_{\odot}\right)^{-1/4}\,\mbox{Hz}$; beyond the
detection level of LISA or LIGO for masses $m\gtrsim 10$~kg. Even if
future experiments could reach such an ultralow frequency bandwith,
the effect of both gravitational radiation and dark energy accretion
is negligible, thus also excluding electromagnetic detection of
changes in the period of the system.

If dark energy accretion is the dominating effect, we can solve
for radius and the angular frequency as a function of time,
\begin{equation}
r(t)= r_0\left(1-\frac{t}{\tau}\right)^2\, ,
\label{eq:rt}
\end{equation}
and 
\begin{equation}
\omega(t)= \omega_0\left(1-\frac{t}{\tau}\right)^{-7/2}\, ,
\label{eq:omegat}
\end{equation}
where again $\tau$ is given in Eq.~(\ref{eq:tau}), corresponding to $\tau\sim 10^{26}$ years for a typical system of supermassive
black hole binaries. 

The time to coalescence due to gravitational radiation for equal
masses $m=10^{8}m_{\odot}$ with separation $r=10^{19}$~m is
$\tau\sim10^{24}$~years, also far beyond the Hubble time. Other
effects, such as dynamical friction with surrounding stars and gas
will undoubtly dominate, and are needed to bring the system to a
separation where gravitational radiation becomes important (roughly
$10^{3}$ Schwarzschild radii
\cite{maggiore}). To bring the system down to separations of a few
hundreds of a parsec using external mechanisms is known as the "final
parsec problem" \cite{parsec}, and dark energy accretion is clearly
not a driving force in these circumstances.

\section{4. Examples}
Two concrete examples where dark energy accretion is claimed to be of
importance are given in \citet{mersini} which we review and comment on
here. The first one is the radio galaxy 0402+379 \cite{rodriguez}. It
is believed to have two supermassive black holes in its center with a
separation of 7.3 parsecs ($\sim2.2\times10^{17}$~m) and total mass
$1.5\times10^{8}m_{\odot}$. The orbital period is 150~000 years. The
radial change due to gravitational radiation is
$\left|\dot{r}_{gw}\right|\sim10^{-9}\,\mbox{m s}^{-1}$ and the change due to dark
energy accretion is $\left|\dot{r}_{de}\right|\sim10^{-16}\,\mbox{m s}^{-1}$. The
merging time due to gravitational radiation is
$\tau\sim10^{18}$~years. This is an example of a system that will
spend the majority of its lifetime in the separation region
$0.1-10$~parsec unless an external mechanism, such as dynamical
friction, carries angular momentum from the system. In any case, unlike
\citet{mersini}, we find that dark energy accretion will be completely 
negligible in this system, as is evident from the smallness of
$\dot{r}_{de}$.

The next example is the quasar OJ287. This object has been observed
since 1891 and shows a 12 year periodic optical outburst. This is
believed to be the result of the passing of a black hole of mass
$m_{1}\sim10^{8}m_{\odot}$ through the accretion disk which belongs to
a black hole of mass $m_{2}\sim10^{10}m_{\odot}$ \cite{valtonen2}. The
system is believed to coalesce in $\tau\sim10^{4}$ years. To model the
system correctly, post-Newtonian approximation schemes must be
applied. This means that the system can be used for precise testing of
general relativity and the effect of gravitational radiation
\cite{valtonen1}. The system is described by a post-Newtonian
Keplerian orbit with eccentricity $e=0.66$. Thus, we cannot use
Eq.~(\ref{eq:radtot}). To obtain an order of magnitude comparison
between the effect of gravitational radiation and dark energy
accretion, we idealize the system as being in a circular orbit with
equal masses $10^{9}m_{\odot}$ (as is done in \citet{mersini}). If we
take the semi-major axis as the radius of the orbiting black hole, we
have $r\sim10^{15}$~m. This result in a radial change of
$\left|\dot{r}_{gw}\right|\sim10\,\mbox{m s}^{-1}$ and
$\left|\dot{r}_{de}\right|\sim10^{-18}\,\mbox{m s}^{-1}$. Again, we find that the
effect from gravitational radiation is the dominating effect, contrary
to the claims in
\citet{mersini}.

\section{5. Conclusions}
We have investigated the effect of dark energy accretion in binary
supermassive black hole systems. When comparing the mass change due to
accretion from the interstellar medium, dark matter and dark energy,
we find that the effect of dark energy accretion for all practical
purposes is negligible. We also compare the effects from dark energy
accretion and the loss of energy through gravitational wave emission. At the separations between the binary constituents needed for dark
energy accretion to dominate over gravitational radiation, and under
the assumption that the rotation of dark energy in the vicinity of the
binary system is not greatly exceeding that of the binary system, we find the
radial change induced by the accretion to be too small to be
observable. Thus, even in an idealized setting with no gas accretion,
the effect of dark energy accretion during the inspiraling phase does
not have a measurable impact on the dynamics of the system. Doing
local measurements of the equation of state of dark energy, as
proposed in \citet{mersini}, can therefore unfortunately not be
considered possible.

\begin{acknowledgments}
EM acknowledge support for this study by the Swedish Research Council.
\end{acknowledgments}

\bibliography{smbhde}

\end{document}